\documentclass[prb,preprint]{revtex4}
\usepackage{graphicx}
\usepackage{hyperref}
\usepackage{times}

\begin{document} 

\title{A Home Experiment in Elasticity}
\author{
J.\ M.\ Aguirregabiria, 
A.\ Hern\'{a}ndez 
and 
M.\ Rivas 
} 
\affiliation{Theoretical Physics, 
The University of the Basque Country, \\
P.~O.~Box 644,
48080 Bilbao, Spain}

\bigskip

\begin{abstract} 
We analyze a simple problem in elasticity: the \emph{initial} motion of an elastic
bar that after being hanged from an end is suddenly released. In a second
problem a point mass is attached on the top of the bar.
The analytical solutions uncover some unexpected properties, which can be  checked, 
with a digital camera or camcorder,
in an alternative  setup in which a spring is substituted for the bar. 
The theoretical model and the experiments are useful to understand
the similarities and differences between the elastic
properties of bar and spring.
Students can take advantage of the home experiments to improve their understanding of
elastic waves. 
\end{abstract} 

\maketitle


\section{Introduction}\label{sec:intro}

\begin{quotation}
``A thin long elastic bar is vertically suspended from one end by means of a string.
After the equilibrium is attained, the string is suddenly cut. Will all bar points
have the same acceleration \emph{just after the string is cut}? ---asks the teacher.
\\Yes sir ---answers Peter--- according to the equivalence principle all
particles will fall with the same acceleration in an homogeneous gravitational field.
\\But I have read in the solution to problem 79 of B\'ujovtsev's book \cite{bb:bujovtsev}
that although the lower end will have acceleration $g$ the upper end will fall faster
---says Ann.
\\Of course ---stresses Mary---, when we computed the tension in the equilibrium
state we saw that the tension is null in the lower end, so that only gravity
will exert a force on particles there, while the tension is maximum in the
upper end, which will in consequence have higher acceleration.
\\What do you think?''
\end{quotation}

When, some weeks ago, one of us proposed the previous question to our students we did not 
suspect the full interest of the problem. Obviously the answers in the quotation
are all wrong. The equivalence principle only applies to free falling particles and  
in the problem there are also elastic forces. On the other hand, $g$ will be the acceleration
of the center of mass: so it is not possible to have an acceleration equal to $g$
at the lower end and greater values everywhere else, as Ann and Mary think.
Our students found the problem highly unintuitive. It took some time to
convince them that although the tension in the upper end disappears when
the string is cut, it will pass some time before the tension and, as a
consequence of Hooke's law, the deformation change in other points. The
change of state will propagate along the bar in the form of an elastic
wave, so that \emph{initially} the lower end does not move at all. 

To help them understand the phenomenon we devised a simple experiment.
We will show below that the theoretical analysis in the case of
a metallic bar is elementary, but the deformation would be too small and change too
fast to be seen, except maybe with some sophisticated experimental
setup. Instead we used  a plastic spring \cite{bb:slinky}, which
when stretched behaves much like the elastic bar, but
have elastic properties that change completely when the loops are in contact with each other.
In the spring deformations are large and change slowly enough to be recorded with a
standard digital camera, since most of them have the ability to record short
video sequences. Although the result of our first try \cite{bb:first} 
was not of good quality, one
could easily see that the lower end did not start falling until the
deformation change reached it.

\begin{figure}
\begin{center}
\includegraphics[height=.7\textwidth]{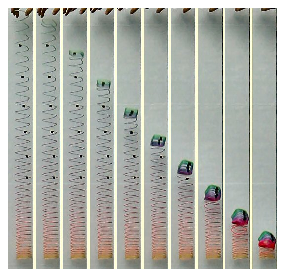}
\end{center}
\caption{Some consecutive frames from the video sequence \cite{bb:magic}.\label{fig:first}} 
\vspace{3cm}
\begin{center}
\end{center}
\end{figure}

One could then think that if the center of mass
moves with acceleration $g$ while the points at the lower end are still at rest, the
upper points must move with high accelerations. We will see in the following that
this is not the case with the bar: what happens is something simpler but (probably) less intuitive.
In fact,  one can go
beyond the qualitative analysis and compute the full evolution (from the
string being cut to the deformation change reaching the lower end) with a
minimal mathematical apparatus, by using concepts available to students
of introductory physics courses. The analysis of Section~\ref{sec:sol1} 
predicts that after the
deformation change has reached a point it will start moving with a velocity
independent of position and time, i.e., without acceleration. This unexpected
result is a consequence of having assumed that the force exerted by the string
vanishes instantaneously, but one could expect it to be a good approximation
in appropriate cases. To check the theoretical prediction we went back to our 
video, extracted some consecutive frames with a free tool \cite{bb:virtualdub} 
and get something like Figure~\ref{fig:first}, where one can
see that the upper end moves with a more or less constant velocity, while the lower 
end remains at rest until the elastic wave reaches this point. 

However, as discussed in Section~\ref{sec:exper},
except for the first frames in Figure~\ref{fig:first}, there was not good 
agreement with the theoretical analysis because the upper coils quickly became
completely compressed and touched one another. 
The reason can be understood by using the theoretical model: 
matter quickly moves faster than the elastic wave and the dynamical problem changes
completely (this new problem was discussed by Calkin~\cite{bb:calkin}).

We then realized that we could get better agreement between the theory of elastic waves
and the experiment with the spring by modifying the problem
by attaching a mass on top of bar and spring. As described in Section~\ref{sec:sol2},
however small the mass is, some properties of the solution for the bar 
change qualitatively, and not only quantitatively. This in turn allows to apply  
for longer spans of time
the theoretical model for the bar to the experimental setup with the spring.

\section{The falling bar}\label{sec:sol1}

\begin{figure}
\begin{center}
\includegraphics[height=.7\textwidth]{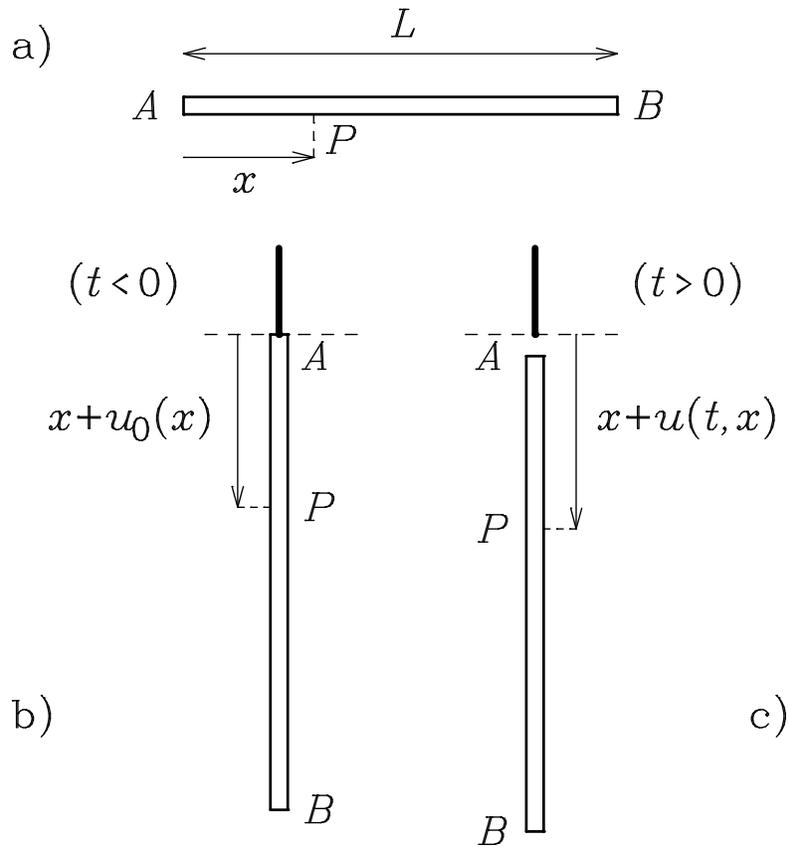}
\end{center}
\caption{Elastic bar a) without strain, b) hanging and  c) released.\label{fig:bar}} 
\vspace{3cm}
\begin{center}
\end{center}
\end{figure}

In the following we will label each 
point $P$ with the distance $x$ measured from end $A$ when the bar is not strained, 
as shown in Figure~\ref{fig:bar}~a). 

\subsection{Initial equilibrium}\label{sec:initial}
For times $t<0$ the bar is hung
as shown in Figure~\ref{fig:bar}~b), so that the distance
$AP$ is now $x+u_0(x)$ in terms of the deformation field $u_0$. The tension $\tau_0$ is
readily computed by writing the equilibrium condition of $PB$ and using the
fact that tension vanishes at $B$: $\tau_0(L)=0$. One gets
\begin{equation}\label{eq:tau1}
\tau_0(x)=\rho g (L-x),
\end{equation}
where $\rho$ is the mass density. Hooke's law and the boundary condition $u_0(0)=0$
allow computing the deformation:
\begin{equation}\label{eq:u1}
\tau_0=E\,\frac{\partial u_0}{\partial x}\quad\Longrightarrow\quad u_0(x)=\frac{g}{2c^2}\left(2Lx-x^2\right),
\end{equation}
where $E$ is Young's modulus and $c\equiv\sqrt{E/\rho}$  the speed of sound in the bar.
We are assuming here the unitary deformation $\partial u_0/\partial x$ is small enough
to satisfy Hooke's law, which implies $gL/c^2$ is small.

\subsection{Initial dynamics}
At $t=0$
the string is cut so that the tension at $A$ disappears instantaneously,  and
a discontinuity starts propagating along the bar with velocity $c$. It is easy
to use Newton's second law and Hooke's law to find the wave equation satisfied by the
deformation field $u(t,x)$, which gives the position of $P$ at any time, as depicted
in Figure~\ref{fig:bar}~c):
\begin{equation}
\frac{\partial^2 u}{\partial t^2}=c^2\frac{\partial^2 u}{\partial x^2}+g.
\end{equation}
However, probably it will be easier for students to understand the calculation in
a reference frame falling with the center of mass, where the deformation field will be
\begin{equation}\label{eq:galileo}
u^\ast= u-\frac12gt^2.
\end{equation}
In such a frame the inertial force and weight  cancel each other and the
remaining forces are of elastic origin, so that the longitudinal
evolution equation is the homogeneous wave equation introduced in
elementary physics courses \cite{bb:elem},
\begin{equation}\label{eq:waveeq}
\frac{\partial^2 u^\ast}{\partial t^2}=c^2\frac{\partial^2 u^\ast}{\partial x^2},
\end{equation}
whose solution is early taught to students in d'Alembert's form, i.e., as 
a superposition of two waves
propagating in opposite directions~\cite{bb:elem}:
\begin{equation}\label{eq:fplug}
u^\ast(t,x)=f^\ast(x-ct)+h^\ast(x+ct).
\end{equation}
(Of course, this is the standard way of solving inhomogeneous linear equations,
but the reasoning here is more physical and can be presented to students before
they study differential equations.)

For $x>ct$ the perturbation has not reached the point, so that we have $u^\ast=u^\ast_0$. 
In consequence we seek a piecewise solution as follows:
\begin{equation}
u^\ast(t,x)=\left\{
\begin{array}{ll}
\displaystyle u_0(x)-\frac12gt^2,&x>ct;\\
\displaystyle f(x-ct)+h(x+ct),&x<ct<L.
\end{array}\right.
\end{equation}
For $x<ct$ we have a solution of (\ref{eq:waveeq}) in terms of two functions, $f$ and $h$, 
which is easily computed by using the following two physical conditions:
\begin{itemize}
\item The bar does not break so that  $u$ is continuous at the wavefront $x=ct$:
\begin{equation}\label{eq:h}
-\frac{g}{4c^2}2x(2x-2L)=f(0)+h(2x)\quad\Longrightarrow\quad h(x)=\frac{g}{4c^2}x(2L-x)-f(0).
\end{equation}
Since only the combination $f(x-ct)+h(x+ct)$ will appear in the solution, 
there is no restriction in taking $f(0)=0$.
\item In this reference frame the center of mass is at rest, so that its velocity is:
\begin{eqnarray}
0&=&\frac1L\int_0^L{\frac{\partial u^\ast(t,x)}{\partial t}\,dx}
\nonumber\\
&=&\frac1L\int_{ct}^L{\left(-gt\right)\,dx}+\frac cL\int_0^{ct}{\left[h'(x+ct)-f'(x-ct)\right]\,dx}
\nonumber\\
&=&\frac cL\,f(-ct)-\frac{g}{4L}\,t (2L-ct). \label{eq:momnul}
\end{eqnarray}
\end{itemize}
In consequence we have
\begin{equation}
 f(x)=-\frac{g}{4c^2}x(2L+x).
\end{equation}
Remembering (\ref{eq:galileo}), we get our main result in the laboratory frame:
\begin{equation}\label{eq:main}
u(t,x)=\frac g{2c^2}\left\{
\begin{array}{ll}
2Lx-x^2,&x>ct;\\
2Lct-x^2,&x<ct<L.
\end{array}
\right.
\end{equation}

\subsection{Properties of the solution}\label{sec:properties}

The stress is then 
\begin{equation}\label{eq:stress}
\tau(t,x)=E\,\frac{\partial u}{\partial x}=\rho g\,\left\{
\begin{array}{ll}
L-x,&x>ct;\\
-x,&x<ct<L.
\end{array}
\right.
\end{equation}
We have, as expected, $\tau(t,0)=\tau(t,L)=0$ for all $0<t<L/c$. On the other hand, the tension
is discontinuous and 
becomes compression at the wavefront, and solution (\ref{eq:stress}) at the lower end $x=L$
goes to $-\rho gL$ as $t\to L/c$. This proves there must appear a reflected wave to make sure
$\tau(t,L)=0$ always; but we are not interested here in this reflected wave, for it cannot be seen 
in our home experiment.

The surprise arises when one computes the velocity of each point:
\begin{equation}\label{eq:velocity}
\frac{\partial u}{\partial t}=\frac{gL}c\,\left\{
\begin{array}{ll}
0,&x>ct;\\
1,&x<ct<L.
\end{array}
\right.
\end{equation}
All points outside the wavefront move without acceleration; but more
and more points quit rest and start moving with velocity $gL/c$ (which,
as stressed after (\ref{eq:u1}), is smaller than $c$)
as time
increases, so that the center of mass moves with increasing velocity
$gt$. Put in other way, at the wavefront  $x=ct$ the velocity is discontinuous and,
in consequence, the acceleration infinite. For students knowing
Dirac's delta function ---which can be easily introduced to physics students \cite{bb:delta}---
one can write the acceleration in the following form:
\begin{equation}
\frac{\partial^2 u}{\partial t^2}=gL\,\delta(x-ct),\qquad(t<L/c).
\end{equation}
We see now that the answer to the question proposed at the beginning of the introduction
is that, in the limit in which the string is cut instantaneously, initially
all points move without acceleration, except for those points lying at the wavefront,
which have infinite acceleration.
This problem illustrates a rather unusual way for a system of particles
to gain  more and more linear momentum under an external force: it is
enough to keep putting particles in motion, although all velocities are equal
and, more strikingly, constant.

\section{The first experiment}\label{sec:exper}

The theoretical analysis of the problem with an elastic bar can be carried out
with elementary physics; in fact, it is enough
to use the first example of wave equation discussed in some textbooks \cite{bb:waves}. 
In our home experiment
we needed bigger deformations and slower propagation velocities, so we
considered using instead a spring. The complete study of the latter is more difficult:
deformations are no longer small, the elastic properties when stretched 
and under compression are completely different, when hung from an end 
it stretches but also develops non negligible torsion which changes when
moving, transversal motion is likely to appear,
and so on. However, one
could expect that at least some qualitative results (such
as the points starting moving progressively with constant velocity) 
would be the same as in the elastic bar.

Instead of an elastic bar we released a
colorful plastic spring \cite{bb:slinky} with some black tags stuck
every third loop. We used a digital camera \cite{bb:camera} to shoot a short video sequence 
at 30 frames per second. The resulting animation is displayed (at two different
speeds) in a web  page \cite{bb:magic}.  One can clearly see there that
the tags and the lower end remain at rest for a while. To further explore the process, we used  a tool
\cite{bb:virtualdub} to extract some consecutive frames, which are
displayed in Figure~\ref{fig:first}. 

\begin{figure}
\begin{center}
\includegraphics[height=.7\textwidth]{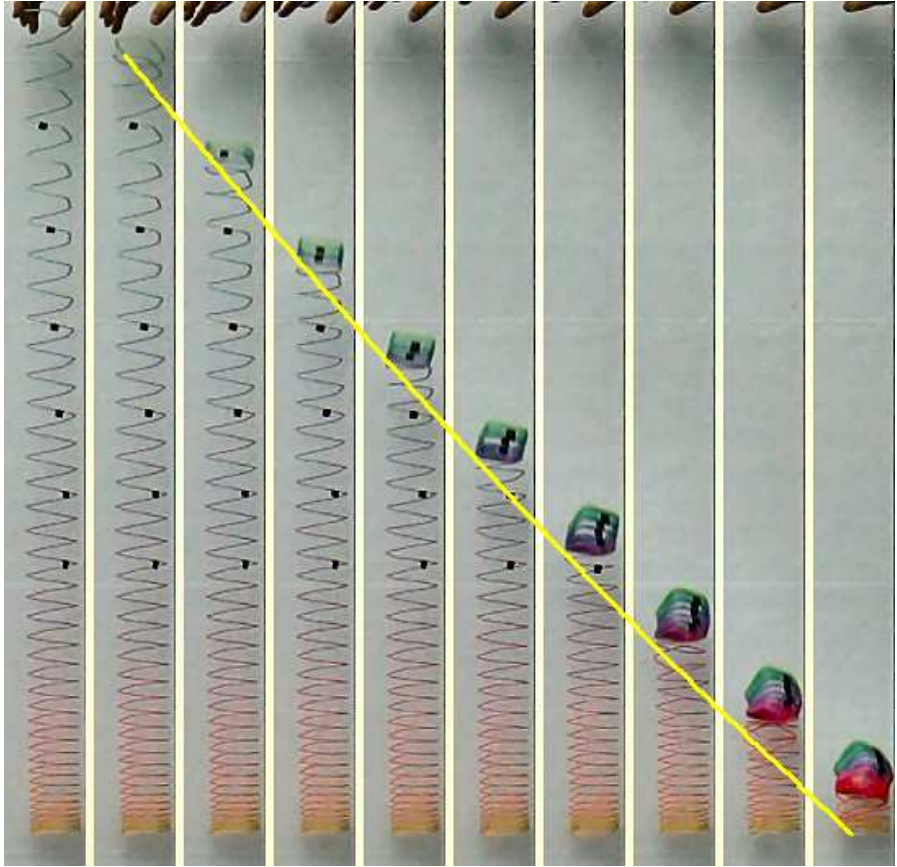}
\end{center}
\caption{Same as Figure~\ref{fig:first} along with the solution by Calkin~\cite{bb:calkin}.\label{fig:fit}} 
\vspace{3cm}
\begin{center}
\end{center}
\end{figure}

At first sight one might conclude that the  tagged points start moving all with the same
constant velocity only when the elastic wave reaches them, but a simple
calculation shows disagreement between theory and experiment: the upper coils
quickly become completely compressed. 
Initially the solution computed for the bar is approximately
valid for the spring ---and was computed in  this context, by means of a
somewhat
less elementary mathematical method by Calkin~\cite{bb:calkin} (see also Cushing~\cite{bb:cushing})---: one
only has to replace $c$ by $L\sqrt{k/m}$, where $m$ and $k$ are
respectively the mass and the elastic constant of the spring. 
But, unlike in a metallic bar, in a
soft spring the velocity $gL/c$ of the coils above the frontwave $x=ct$
quickly become bigger that the velocity of the latter:
\begin{equation}\label{eq:velct}
\frac d{dt}\left[ct+u(t,ct)\right]=\frac{gL}c+c-gt.
\end{equation}
From $t=c/g$ on (in fact, a bit earlier due to the finite thickness of the coils) 
an increasing number of upper coils touch one another 
and fall over the coils below before there is time for the tension to
change there. We have then a kind of matter wave that moves faster than the elastic
wave created when the spring was released. The dynamical problem is then completely different and was 
analyzed by  Calkin~\cite{bb:calkin}.
One can check his solution in our case 
($l_0\approx l_1\approx6.5\mbox{ cm}$, $\xi_1\approx1$, $k/m\approx4\mbox{ s}^{-2}$, in his notation):
good numerical agreement is shown in Figure~\ref{fig:fit}.
This may be an interesting problem in dynamics of systems of particles, but we were
more interested in elasticity, so that we turned to the study of the problem in Section~\ref{sec:sol2}.

We can see that the solution (\ref{eq:main}) breaks at $t=c/g$ (provided it is less than $L/c$, 
which would never happen for an actual metallic bar) from another point of view. 
At that moment 
\begin{equation}\label{eq:cross}
\frac{\partial}{\partial t}\left[x+u(t,x)\right]=1-\frac {gx}{c^2}
\end{equation}
becomes negative at the wavefront $x=ct$, which is clearly impossible, for it would mean
an inversion of the spatial order of coils.

\section{The falling bar with an attached pointlike mass}\label{sec:sol2}

Let us consider again the bar of Figure~\ref{fig:bar} but
let us assume in the upper end $A$ there is attached a pointlike
mass $M=\mu m$. The analysis of Section~\ref{sec:sol1} only changes
from (\ref{eq:momnul}) on. In the latter we now have to
take into account the contribution from $M$:
\begin{equation}
\frac1L\int_0^L{\frac{\partial u^\ast(t,x)}{\partial t}\,dx}+\mu\frac{\partial u^\ast}{\partial t}(t,0)=0,
\end{equation}
which leads to the condition
\begin{equation}
\mu Lf'(x)-f(x)=\frac{g}{4c^2}\left[2\mu L^2+2(1+\mu)Lx+x^2\right].
\end{equation}
This is a first-order linear equation with constant coefficients 
which can be solved, along with the initial condition $f(0)=0$,
 in a number of easy ways (integrating with respect to $x$ after multiplying it with $e^{-x/\mu L}$ or
 by using computer algebra, for instance) to give
 \begin{equation}
f(x)=-\frac g{4c^2}\left[x(2L+x)-2q_\mu(x)\right],\qquad q_\mu(x)\equiv2\mu L\left[(1+\mu)L\left(e^{x/\mu L}-1\right)-x\right].
 \end{equation}

In consequence,  the deformation field in the laboratory frame is
\begin{equation}\label{eq:main2}
u(t,x)=\frac g{2c^2}\left\{
\begin{array}{ll}
2Lx-x^2,&x>ct;\\
2Lct-x^2+q_\mu(x-ct),&x<ct<L,
\end{array}
\right.
\end{equation}
which reduces to  (\ref{eq:main}) in the limit $\mu\to0$, since $\lim_{\mu\to0}q_\mu(x-ct)=0$ for $x<ct$.
The velocity 
\begin{equation}\label{eq:velocity2}
\frac{\partial u}{\partial t}=\frac{g(1+\mu)L}c\left(1-e^{(x-ct)/\mu L}\right)\,\left\{
\begin{array}{ll}
0,&x>ct;\\
1,&x<ct<L
\end{array}
\right.
\end{equation}
is now, unlike in (\ref{eq:velocity}), continuous through the wavefront $x=ct$
for any mass ratio $\mu>0$. 

This qualitatively different behavior may seem at first counterintuitive,
but there is a clear physical reason for it: now the stress 
\begin{equation}\label{eq:stress2}
\tau(t,x)=\rho g\,\left\{
\begin{array}{ll}
L-x,&x>ct;\\
-\mu L-x+(1+\mu)L e^{(x-ct)/\mu L},&x<ct<L
\end{array}
\right.
\end{equation}
does not becomes instantaneously null at $t=0$ and $x=0$, 
but retains its previous value, $\tau(0,0)=\rho gL$, 
because of the attached mass, however small it is.
For the same reason the points above the frontwave are now 
accelerated:
\begin{equation}
\frac{\partial^2 u}{\partial t^2}=g\,\frac{1+\mu}\mu\, e^{(x-ct)/\mu L},\quad\mbox{for } x<ct<L.
\end{equation}

Although now the velocity at the wavefront is zero, the solution would break at the moment
\begin{equation}
t=\frac{\mu L}c\log\frac{(1+\mu)L}{\mu L+x-c^2/g}+\frac xc,\quad(x<ct<L),
\end{equation}
provided it is real and less than $L/c$, 
as one can see by repeating the calculation in (\ref{eq:cross}). One can check that in 
the case of our spring this value increases with $\mu$, so that one would expect our analytical
solution~(\ref{eq:main2}) to be valid for longer intervals with heavier masses.

Notice that  when the elastic wave reaches a point its acceleration is $g(1+\mu)/\mu>g$,
and the same happens initially at the upper end $A$.

\section{The second and third experiments}\label{sec:second}

\begin{figure}
\begin{center}
\includegraphics[height=.7\textwidth]{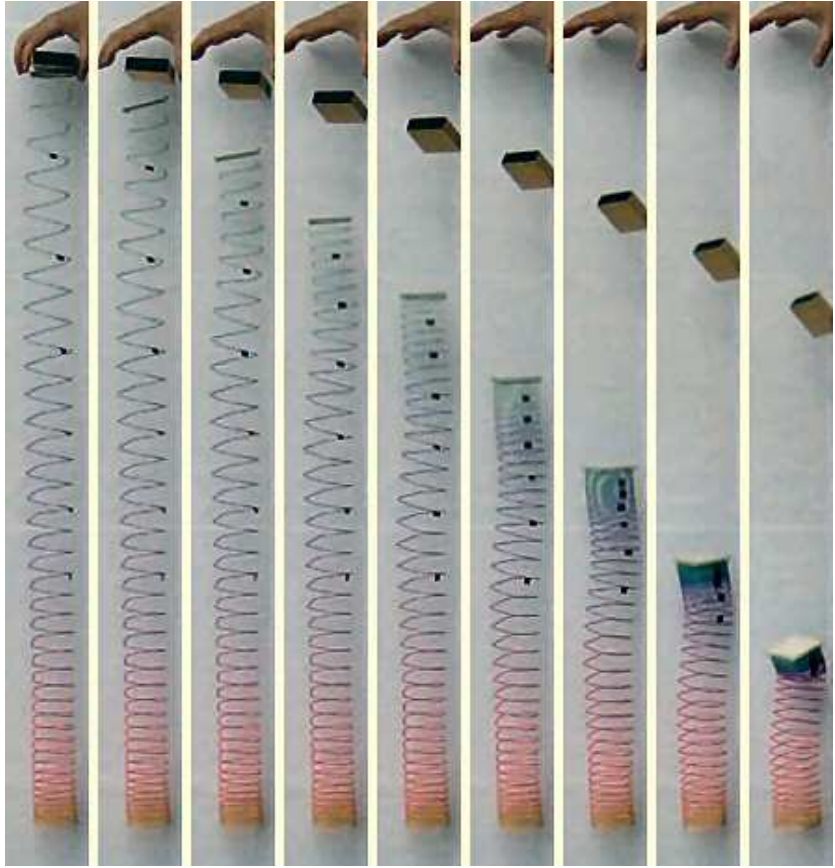}
\end{center}
\caption{Same as Figure~\ref{fig:first} with a thin slab fixed at the upper end and a block on top of it.\label{fig:slab}} 
\vspace{3cm}
\begin{center}
\end{center}
\end{figure}

This can be clearly seen in our second experiment \cite{bb:magic2} where
a thin wooden slab is fixed at the top end $A$. A thicker block is then put on top of it.
When the spring is released, the acceleration of $A$ is greater than that of the block, which
immediately splits up and follows the familiar free fall  trajectory,
as displayed in the consecutive frames of Figure~\ref{fig:slab}. One can see there
that the lightweight mass ($\mu\approx 0.22$) is enough to make sure the upper coils are also stretched
for a while and, in consequence,
to avoid our continuous theoretical model break early because of the finite
thickness of the coils and the fact that the  elastic properties of the spring change completely when
the coils touch one another. 

\begin{figure}
\begin{center}
\includegraphics[height=.7\textwidth]{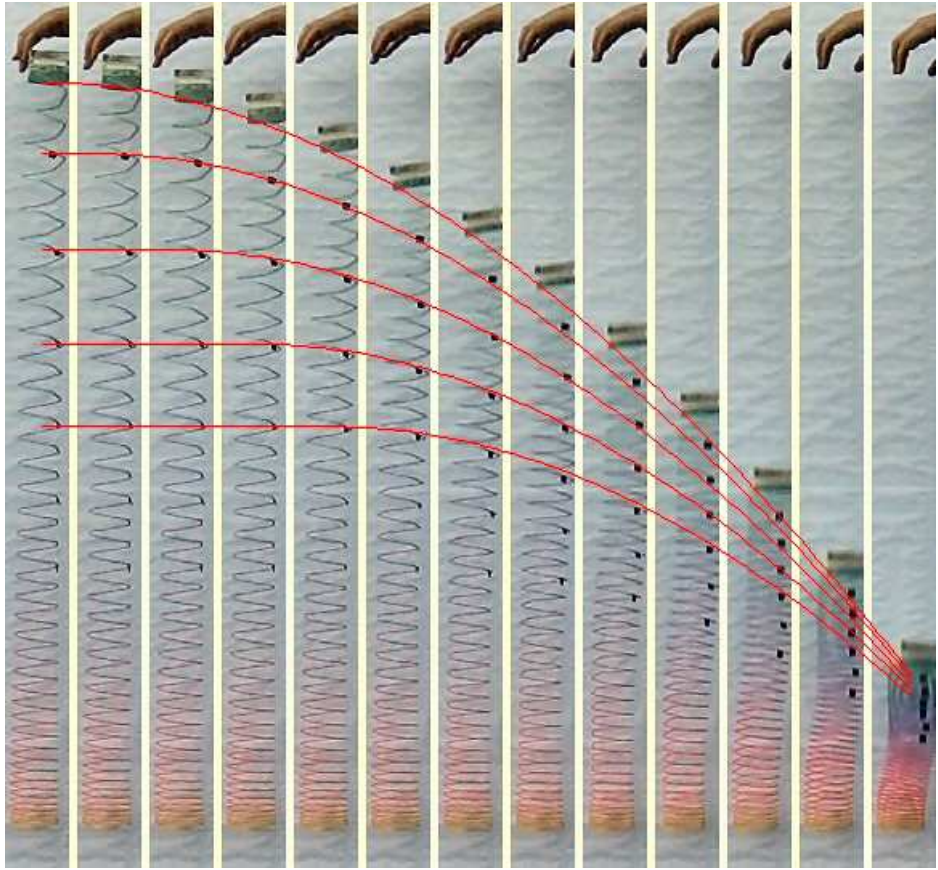}
\end{center}
\caption{Same as Figure~\ref{fig:first} with a wooden block ($\mu\approx2.35$) fixed at the top end.\label{fig:block}} 
\vspace{3cm}
\begin{center}
\end{center}
\end{figure}

This can be seen even better in our third
experiment \cite{bb:magic3}, where the thin slab is replaced by a thicker block ($\mu\approx 2.35$).
One can clearly see in Figure~\ref{fig:block} that each black tag starts moving only when the
stretching begins to change there, i.e., when the elastic waves reaches it.
In the same figure we have plotted the trajectories of the top most coil and the first three tags
as computed with solution~(\ref{eq:main2}): one gets good agreement, 
despite the amateurish experiment and the differences between bar and spring.

\section{Conclusions}\label{sec:concl}

We have analyzed, by using elementary physical concepts, a couple of
problems in elasticity with results difficult to anticipate by pure
intuition. To help students understand elastic waves one can easily
check  the most striking aspects of the analytical results in an alternative
 setup, in which the elastic bar is substituted
by a spring, which allows far  bigger deformations and slower wave
propagation. The similarities and differences between bar and spring
can be used in an illustrative discussion in introductory physics courses.

Since one only needs a digital camera (or a video camera) and some
freeware to process the video sequences, the experiments can be easily
performed in the classroom and repeated at
home by the interested student.

One can take advantage of the widespread availability of digital cameras 
to visually check the solution of other problems in mechanics.
Let us only mention a well-known one: the free end of an articulated arm
released from an angle below $\arcsin 1/\sqrt3\approx35^\circ$   is always accelerated faster than 
$g$~\cite{bb:bolatxoa}.

\acknowledgments
This work was supported by The University of the Basque Country
(Research Grant~9/UPV00172.310-14456/2002).


\clearpage

%
%
%
%
%

\clearpage

\clearpage
\clearpage
\clearpage
\clearpage

\end{document}